\documentclass[journal=nalefd,manuscript=article]{achemso}

\usepackage{achemso}
\usepackage{chemformula} 
\usepackage{graphicx} 
\usepackage{amsfonts}
\usepackage{algorithm}
\usepackage{algpseudocode}
\usepackage{subcaption}
\usepackage{float}

\title{Polarization Domain Mapping From 4D-STEM Using Deep Learning}

\author{Fintan G. Hardy}
\affiliation{Department of Materials, London Centre of Nanotechnology, Imperial Henry Royce Institute, Imperial College London, SW7 2AZ, UK}

\author{Sinéad M. Griffin}
\affiliation{Materials Sciences Division, Molecular Foundry, Lawrence Berkeley National Laboratory, Berkeley, CA 94720, USA}

\author{Mariana Palos}
\affiliation{Department of Materials, London Centre of Nanotechnology, Imperial Henry Royce Institute, Imperial College London, SW7 2AZ, UK}

\author{Yaqi Li}
\affiliation{Department of Materials, London Centre of Nanotechnology, Imperial Henry Royce Institute, Imperial College London, SW7 2AZ, UK}

\author{Geri Topore}
\affiliation{Department of Materials, London Centre of Nanotechnology, Imperial Henry Royce Institute, Imperial College London, SW7 2AZ, UK}

\author{Aron Walsh}
\affiliation{Department of Materials, London Centre of Nanotechnology, Imperial Henry Royce Institute, Imperial College London, SW7 2AZ, UK}
\email{a.walsh@imperial.ac.uk}

\author{Michele Shelly Conroy}
\affiliation{Department of Materials, London Centre of Nanotechnology, Imperial Henry Royce Institute, Imperial College London, SW7 2AZ, UK}
\email{mconroy@imperial.ac.uk}

\usepackage[inkscapelatex=false]{svg}
\usepackage{graphicx}
\usepackage{dcolumn}
\usepackage{bm}
\usepackage[version=4]{mhchem}

\begin{document}

\begin{abstract}

Polarization in ferroelectric domains arises from atomic-scale structural variations that govern macroscopic functionalities. The interfaces between these domains--known as domain walls--host distinct physical responses, making their identification and control critical. Four-dimensional scanning transmission electron microscopy (4D-STEM) enables simultaneous acquisition of real- and reciprocal-space information at the atomic scale, offering a powerful platform for domain mapping. However, conventional analyses rely on computationally intensive processing and manual interpretation, which are time-consuming and prone to misalignment and diffraction artifacts. Here, we present a convolutional neural network that, with minimal training, classifies polarization directions from diffraction data and segments domains in real space. We further introduce an adaptive sampling strategy that prioritizes images from domain wall regions, reducing the number of training images required while improving accuracy and interpretability. We demonstrate this approach for domain mapping in ferroelectric boracite, \ce{Cu3B7O13Cl}.

\end{abstract}

\maketitle


\section{\label{sec:level1}Introduction}

The ability to map polarization of domains and domain walls is essential for understanding functional properties in ferroelectric materials, where subtle structural variations influence electronic and mechanical behaviour. \cite{jia2008atomic}. 
Advances in electron microscopy techniques, such as four-dimensional scanning transmission electron microscopy (4D-STEM), provide a powerful means to probe these domains by simultaneously capturing real and reciprocal space information. \cite{ophus2019four} \cite{das2019observation} However, extracting meaningful insights from such datasets remains computationally intensive, particularly for complex materials where conventional analysis methods rely on manual interpretation. Artificial intelligence (AI) methods have facilitated a revolution in a vast array of other microscopy domains \cite{spurgeonDatadrivenNextgenerationTransmission2021c}, including the use of autoencoder based methods to de-noise images \cite{sadriUnsupervisedDeepDenoising2024, sangRevolvingScanningTransmission2014}, classification and segmentation of STEM images \cite{robertsDeepLearningSemantic2019, ziatdinovDeepLearningAtomically2017a, leeSTEMImageAnalysis2022}, and the probing of latent spaces in variational autoencoders to extract features from 4D-STEM datasets \cite{oxleyProbingAtomicscaleSymmetry2021}.

\begin{figure}
    \centering
    \includegraphics[width=8.3cm]{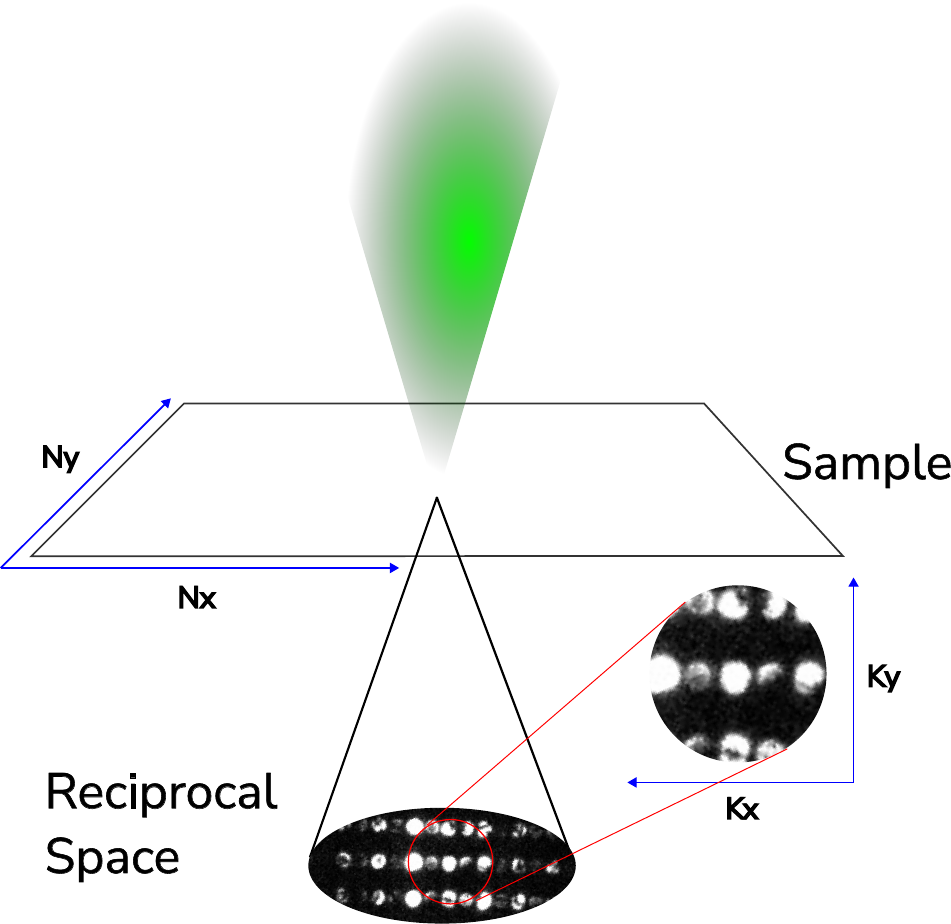}
    \caption{Diagram illustrating a two-dimensional convergent-beam electron diffraction pattern of a \ce{Cu3B7O13Cl} sample. The inset shows a magnified ghost disk. Real space ($N_{x}, N_{y}$) and reciprocal space ($K_{x}, K_{y}$) are displayed along the dimensions of both the sample and the diffraction image, respectively. The magnified section displays the faint ghost disks present in the diffraction images.}
    \label{fig:diffraction_ghost}
\end{figure}

4D-STEM enables high-resolution structural characterization by recording diffraction patterns at each scanned probe position. This microscopy technique generates a four-dimensional dataset, with two spatial dimensions $(N_x, N_y)$ representing the real-space scan coordinates and two reciprocal-space dimensions $(K_x, K_y)$ corresponding to the diffraction data recorded at each scan position. As a result, the dataset forms a four-dimensional array $(N_x, N_y, N_{kx}, N_{ky})$, where each pixel in the real-space image is associated with a full diffraction pattern (system illustrated in Figure \ref{fig:diffraction_ghost}). The challenge in analyzing polarization domains lies in effectively utilizing this rich cross-modal data set to extract meaningful information about the structural variations that define domains and the boundaries between them. 

The analysis of polarization in ferroelectric materials using 4D-STEM typically involves detecting intensity variations in diffraction disks \cite{nguyenMappingPolarityToroidal2018a} or Kikuchi bands \cite{shaoEmergentChiralityPolar2023a}. While this approach is well-established for classic ABO$_{3}$ ferroelectrics, our study instead focuses on a boracite (M$_3$B$_7$O$_{13}$X) crystal, where symmetry changes introduce additional diffraction features aligned with the polarization direction \cite{conroyInvestigatingFerroelasticityGoverning2022}. These extra diffraction disks--often referred to as `ghost' disks--provide insight into polarization shifts through their displacement in reciprocal space, as demonstrated in Figure~\ref{fig:diffraction_ghost}.

The system we focus on here is the boracite \ce{Cu3B7O13Cl}. Boracite is an improper ferroelectric where polarization is not the main order parameter; instead, it arises from spontaneous strain, through a ferroelastic phase transition from cubic \textit{F\={4}3c} to orthorhombic \textit{Pca2$_1$} space groups, accompanied by uniaxial shear and polarization along \textlangle100\textrangle{} pseudocubic directions.\cite{schnelleMagneticStructuralPhase2015b} 
A characteristic of boracite is its six possible polarization directions, with domain walls and domains responding dynamically to an applied bias. As polarization flips at domain boundaries, neighbouring domains exhibit local variations. Manual 4D-STEM analysis reveals that, at certain angles, these domain walls display an intricate alternating polarization pattern. However, studying these effects at scale is problematic. In situ 4D-STEM experiments generate large amounts of data, often tens of terabytes, placing a high demand on storage and computational resources. Furthermore, the polarization behavior within moving ferroelectric walls does not follow a simple or predictable pattern.

To overcome these limitations, a deep learning model designed to automate the detection of domain walls was developed. This strategy not only streamlines data processing but also mitigates bottlenecks in analysis, ultimately accelerating the investigation of novel ferroelectric materials. In our approach, we pass the full 4D tensor into the model, where the diffraction patterns serve as input features for identifying domain walls. A convolutional neural network (CNN) learns to detect key diffraction-based signatures of polarization changes by analyzing local intensity variations in $(K_{x}, K_{y})$, while simultaneously preserving the spatial context in $(N_{x}, N_{y})$. The trained model then produces a segmentation mask in real space, effectively mapping polarization domains across the sample. By leveraging the expressive power of deep learning, this method enhances the efficiency of 4D-STEM analysis, processing a little over 7000 images in 10 seconds on a standard GPU workstation and enabling rapid and accurate domain segmentation without the need for extensive manual interpretation.

\section{Results \& Discussion}
\subsection{Classifying Polar Domains Using Diffraction Patterns}

\begin{figure}
    \centering
    \includegraphics[width=14cm]{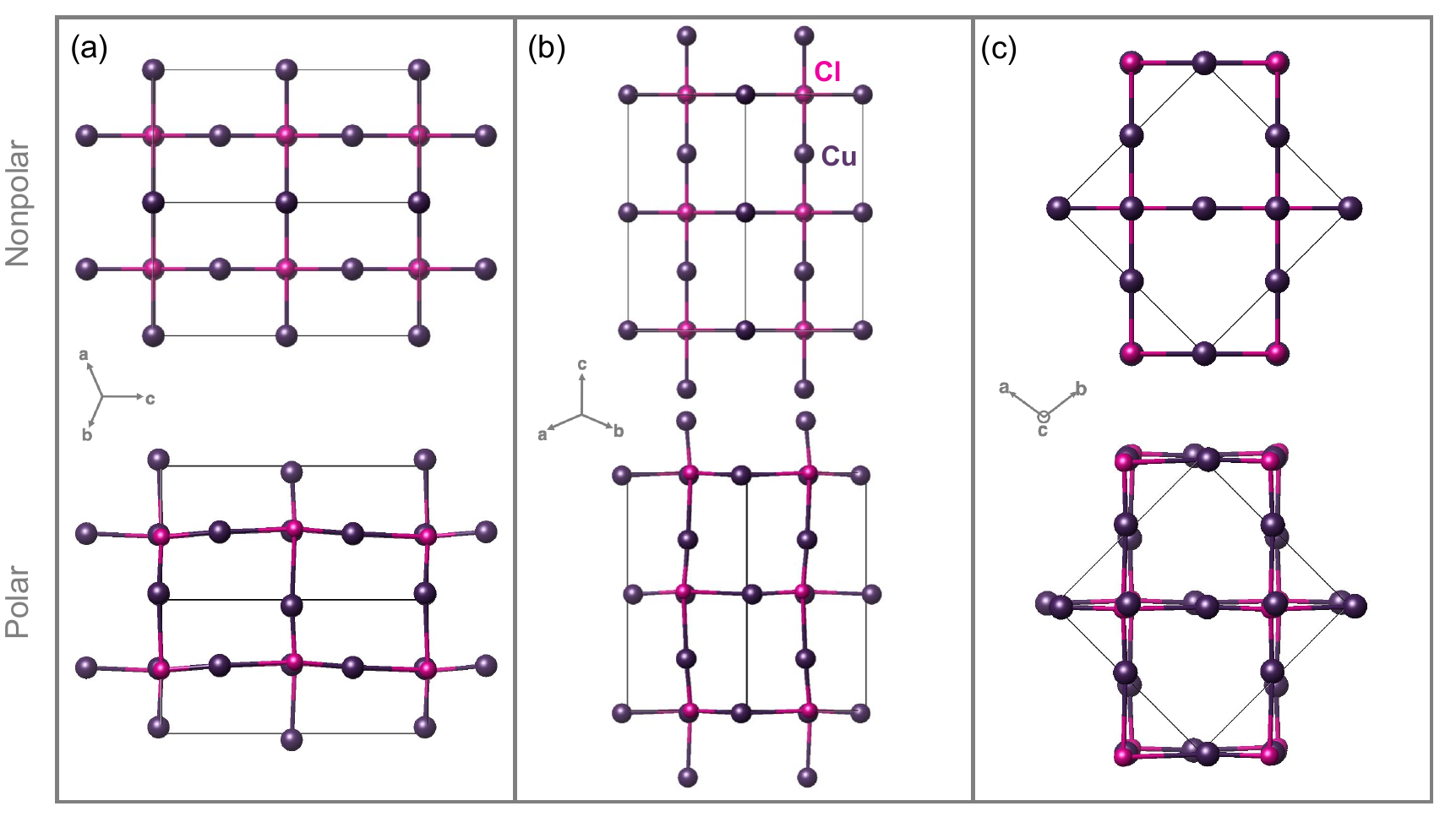}
    \caption{DFT-relaxed structures of Cu$_3$B$_7$O$_{13}$Cl in the non-polar and polar polymorphs, shown along the same three orientations from the STEM measurements. In each orientation pair, the top panel shows the F\={4}3c (non-polar) phase, and the bottom panel shows the Pca2$_1$ (polar) phase. Cu atoms are shown in purple, Cl in pink, while B and O are omitted for clarity. The off-centering magnitude and direction of Cu in the polar phase, relative to the symmetric cubic positions, are evident in each view. These models underlie the simulated STEM images and displacement maps used to train and validate our deep-learning pipeline.}
    \label{fig:DFT}
\end{figure}

In perovskite ferroelectrics, the ability to determine polarization direction from 4D-STEM diffraction patterns relies on identifying subtle intensity variations in specific diffraction features.\cite{nguyenMappingPolarityToroidal2018a}\cite{shaoEmergentChiralityPolar2023a} In boracite, instead of tracking changes in diffraction pattern intensity, the relevant effect is a shift in the polarization vector caused by atomic displacements, leading to crystallographic-plane–dependent extinction and the appearance of additional “ghost” disks. To determine how the atomic displacements influence the polarization vector and the resulting diffraction pattern, we performed Density Functional Theory (DFT) relaxations for both the high-symmetry cubic F$\bar{4}$3c parent phase and the polar Pca2$_1$ ground state of Cu$_3$B$7$O${13}$Cl boracite. Starting from experimental lattice parameters, we fully optimized atomic positions and cell dimensions (see Methods). The resulting equilibrium structures (Fig.~\ref{fig:DFT}) capture the Cu$^{2+}$ off‐centering and Cl$^-$ shifts that generate the net spontaneous polarization, as well as the anti-polar pattern of Cu–Cl displacements in the non-polar phase.  

\begin{figure}
    \centering
    \includegraphics[width=0.5\linewidth]{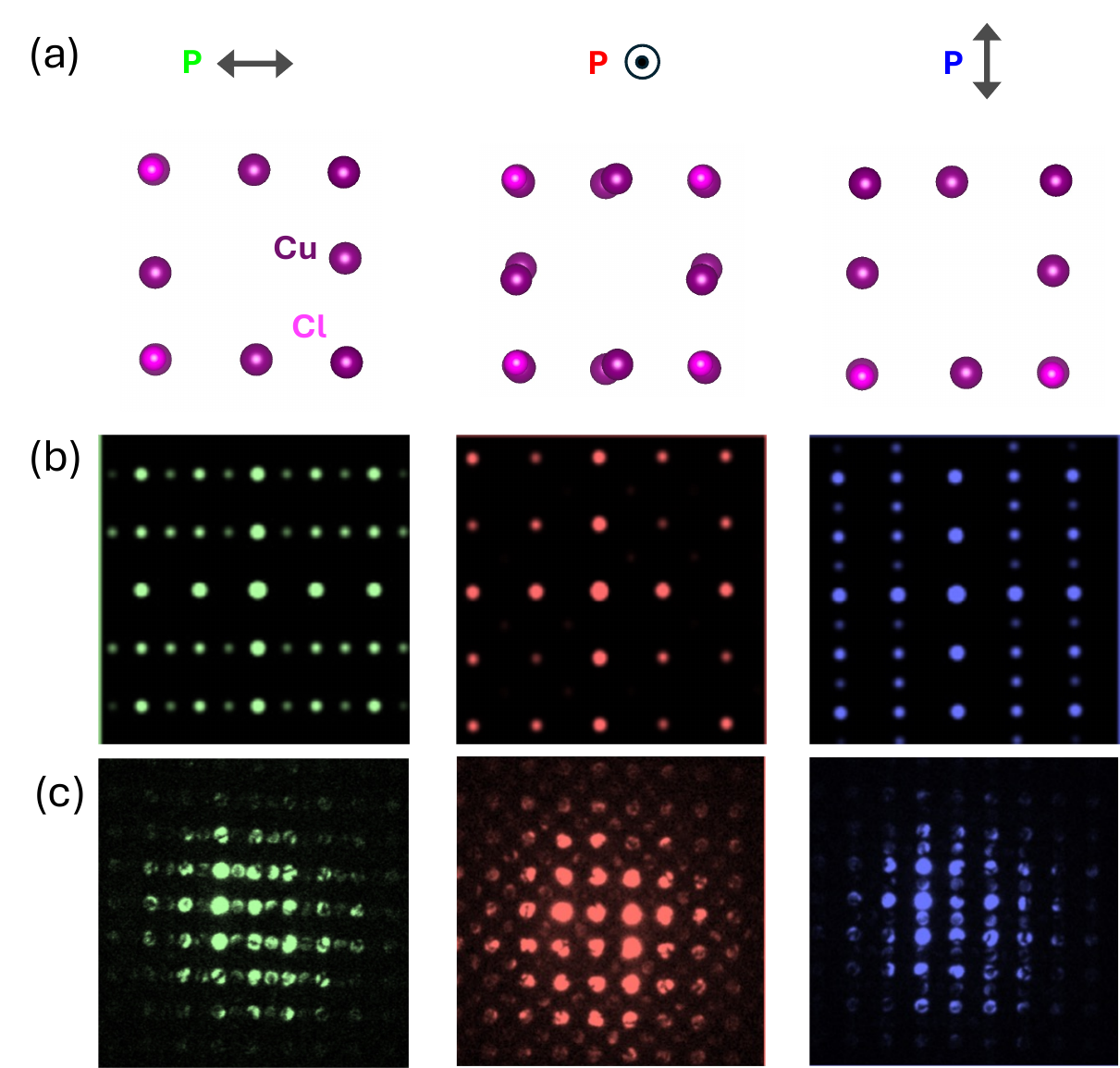}
    \caption{
    a) Boracite sample real-space local crystal structure in horizontal, vertical and out-of-plane polarization directions, 
    b) Simulated diffraction images using directions from a,
    c) Real diffraction images from experiment showing polarization in horizontal, vertical and beam direction polarization.}
    \label{fig:diff-sim}
    
\end{figure}

As shown in Figure~\ref{fig:diff-sim}, we first extracted the main polarization vectors obtained from the DFT calculations and the corresponding structural models. Using these as input, we simulated the diffraction pattern with the SingleCrystal software package. The simulated results reveal distinct shifts in the so-called “ghost” diffraction disk positions (Figure~\ref{fig:diff-sim}b), which can be directly associated with changes in crystal symmetry and, consequently, the underlying polarization states. Importantly, when comparing these simulations with the experimental 4D-STEM diffraction data, we observe additional diffraction disks arising from different domains (Figure~\ref{fig:diff-sim}c). The close agreement between simulated and experimental patterns provides strong evidence that the modifications in the diffraction features are indeed linked to domain-dependent variations in symmetry and polarization.

\begin{figure}
    \centering
    \includegraphics[width=17cm]{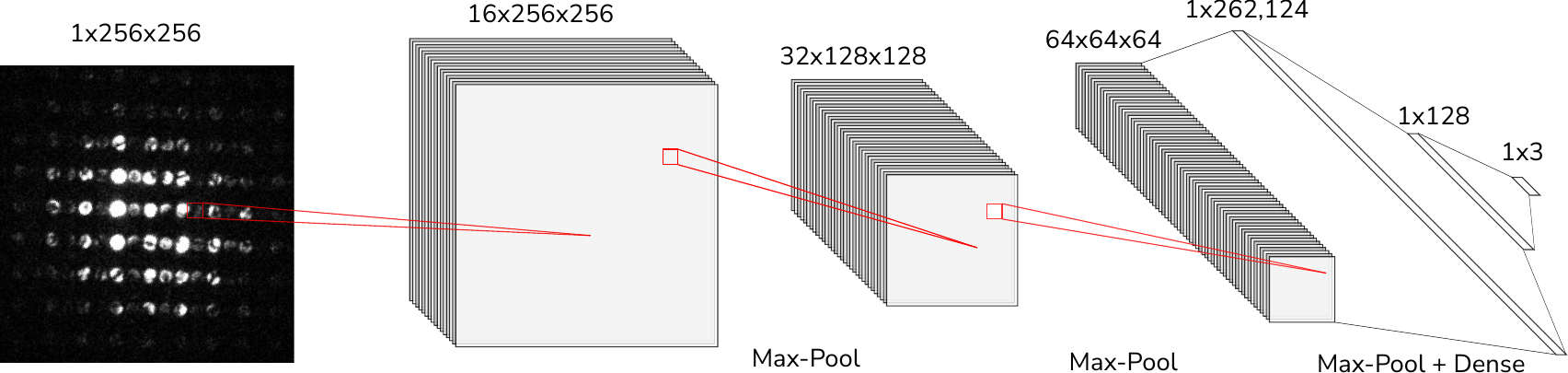}
    \caption{Architecture of our convolution neural network model (\texttt{ghost hunter}) used to predict polarization of individual 4D-STEM images}
    \label{fig:CNN-architecture}
\end{figure}

By analyzing diffraction patterns across the sample, we construct spatial maps of polarization domains. Conventional 4D-STEM data processing can be computationally intensive, as disk-detection methods implemented in py4DSTEM\texttt{py4DSTEM}\cite{savitzkyPy4DSTEMSoftwarePackage2021} or \texttt{pyxem}\cite{francisPyxemScalableMature2023} require repeated fitting or correlation at every probe position, which is impractical for large datasets. To overcome this limitation, we developed a convolutional neural network (CNN) to automate domain classification. The network architecture is shown in Figure \ref{fig:CNN-architecture}, and its output corresponds to the predicted polarization direction: horizontal, vertical, or no polarization. A softmax activation normalizes these outputs to probabilities between 0 and 1, providing classification confidence.

The physical origin of this classification is illustrated in Figure~\ref{fig:diff-sim}. Panel (a) shows representative atomic structures for in-plane horizontal, in-plane vertical, and out-of-plane polarization states. Simulated diffraction patterns in panel (b) reveal symmetry breaking and subtle intensity variations characteristic of each state. Experimental 4D-STEM diffraction images in panel (c) confirm these features, which the CNN exploits to reliably distinguish polarization domains. This integration of experiment, simulation, and machine learning enables high-throughput mapping of domain structures with significantly reduced computational cost.

\begin{figure}[htbp]
    \centering
    \includegraphics[width=8.3cm]{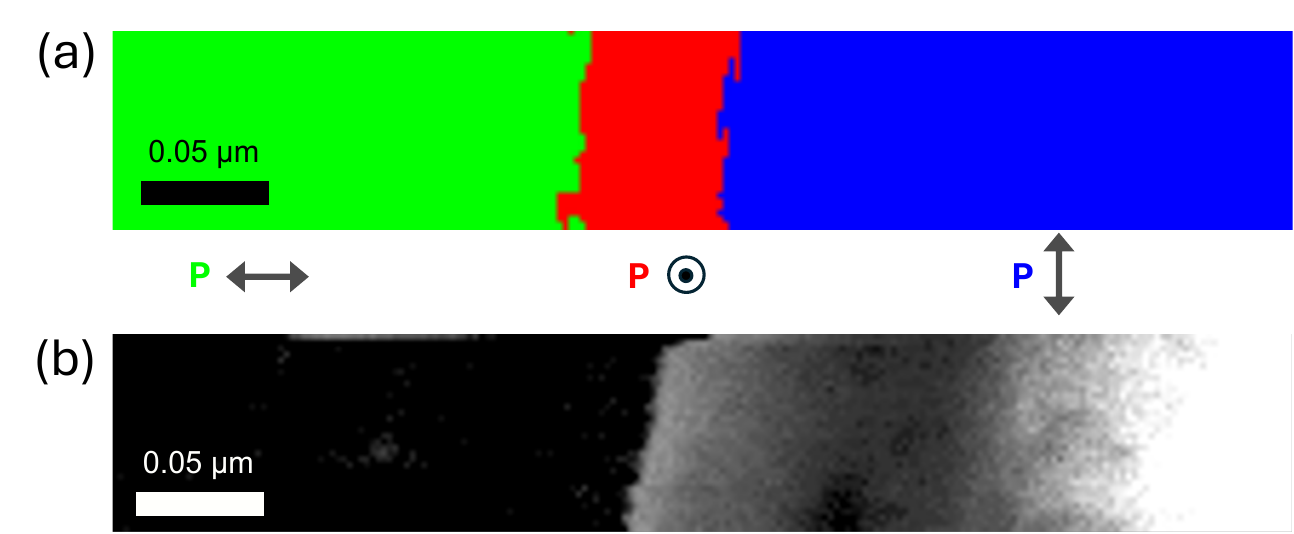}
    \caption{4D-STEM image of \ce{Cu3B7O13Cl}. Real-space segmentation result showing ferroelectric domain structure. Each pixel corresponds to a probe position associated with a diffraction pattern (reciprocal space), and model inference outputs a three-dimensional array representing \textit{no polarization}, \textit{vertical polarization}, and \textit{horizontal polarization}, mapped to RGB values. For example, \([0, 255, 0]\) represents pure green (vertical polarization). 
(a) Segmentation map from convolutional neural network inference. 
(b) Domain wall visualised by manual analysis using the \texttt{py4DSTEM} package. Where darker spots represent more horizontally polarised images and lighter regions represent more vertically polarised images.}
    \label{fig:results_comparison}
\end{figure}

From a total dataset of \( 7,004 \) diffraction images for a \ce{Cu3B7O13Cl} sample, 500 were randomly sampled, and manually labelled based on the direction of the ghost disks, serving as the ground truth for training. The model was trained for 100 epochs using cross-entropy loss, achieving a validation loss of $9\times 10^{-4}$.
During inference, the trained model processed the entire dataset, producing a three-dimensional output for each diffraction pattern corresponding to its predicted polarization state. These values were then mapped to an RGB colour space, with each prediction assigned to its respective spatial position in the real space image, generating a segmentation of the polarization domains. Comparison with manual analysis using \texttt{py4DSTEM} in Figure~\ref{fig:results_comparison} shows how the model follows the same polarization trend with a domain wall positioned in the middle section of the scan.
Here, the region around the domain wall is assigned no polarization (red). This assignment is further confirmed in the labelling patterns observed in Figure ~\ref{fig:binary_search_algorithm}. The absence of this region in the manual analysis is likely due to the experimental conditions in this experiment, where the Bragg peaks are tightly compact, and the method of using offset filters to capture the ghost disks can fail in some cases. 

\subsubsection{Model Performance with Limited Training Data}

To assess the robustness and data efficiency of the CNN model, we conducted experiments using progressively smaller training sets. Instead of training on the full set of 1,000 labelled diffraction images, we tested the model performance with subsets of this dataset, containing 50, 100, and 250 randomly sampled images. The number of training epochs was scaled accordingly to ensure adequate learning while avoiding overfitting, with 200 epochs for the 500-image dataset and up to 400 epochs for the 50-image dataset, where the epoch with the lowest validation loss was selected for each model.

\begin{figure}[htbp]
    \centering
    \includegraphics[width=17cm]{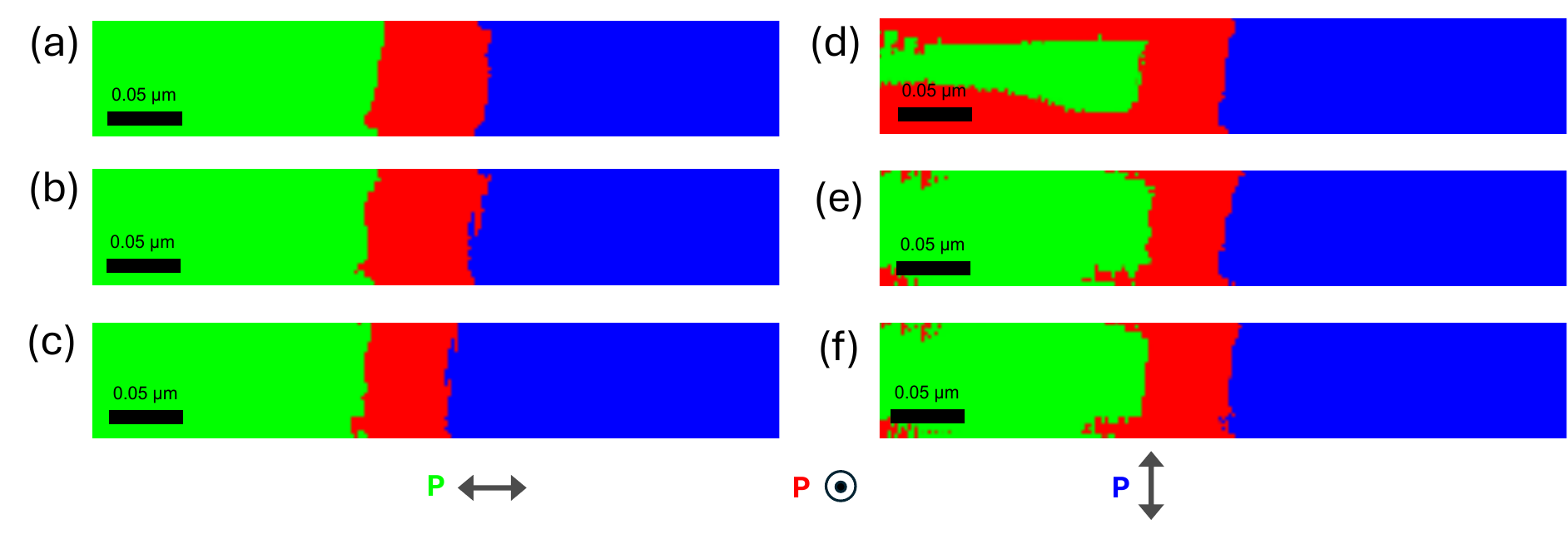}

    \caption{Comparison of CNN segmentation outputs for \ce{Cu3B7O13Cl} 4D-STEM data, trained on different numbers of labelled diffraction patterns (50, 100, 250). 
(a-c) Outputs from models trained on images selected using a binary search sampling algorithm.
(d-f) Outputs from models trained on randomly selected images. $s$ values of 30, 10, and 6 were used for the 50, 100 and 250 image datasets, respectively.}
    \label{fig:labelsize-experiment-comparison}
\end{figure}

Despite the reduced dataset sizes, the model demonstrated an ability to correctly classify polarization directions. As shown for the randomly labelled images d-f in Figure~\ref{fig:labelsize-experiment-comparison}, even with only 50 labelled examples, the CNN identified major polarization regions, albeit with some noise at domain boundaries. As the dataset size increased, the ability of the model to resolve finer details improved, with the 100-image and 250-image models producing results closely matching the full dataset model. Notably, the model was better at detecting neutral polarization regions as more training data was incorporated.

While random sampling shows some success, it is likely that features at domain boundaries, which represent a smaller fraction of the data are undersampled. This leads to an overfit of horizontal and vertical polarization, with an absence of red (non-polar) regions in the right hand-side of Figure \ref{fig:labelsize-experiment-comparison}a. To address this overfitting, a targeted sampling strategy can be introduced to ensure that key diffraction images, particularly those associated with domain walls, are well represented even in smaller datasets. The algorithm consisted of two steps (detailed in the Supporting Information):

\begin{enumerate}
    \item Coarse sampling in a uniform grid pattern across real space $(N_{x}, N_{y})$ with a step size \(s\). The sampled points are given by:
    \[
    \mathbf{p}_i = (N_{xi}, N_{yi}), \quad N_{xi} = i s, \, N_{yj} = j s, \quad i, j \in \mathbb{Z}
    \]
    where \( \mathbf{p}_i \) represents the coordinates of the sampled points in the 2D space. The set of sampled points is:
    \[
    \mathbf{P} = \{ \mathbf{p}_i : 0 \leq N_{xi} \leq X, 0 \leq N_{yi} \leq Y \}
    \]
    where \( X \) and \( Y \) are the dimensions of the image in real space.
    
    \item Adaptive sampling near label boundaries. For each point \( \mathbf{p}_i \in \mathbf{P} \), the algorithm searched its 8 connected neighbours (i.e., the 8 points within a distance less than or equal to the step size \( s \)). Among these neighbours, it identified the closest point \( \mathbf{p}_j \) such that \( L(\mathbf{p}_j) \neq L(\mathbf{p}_i) \). If such a point was found, a new sample was generated at the midpoint:
    \[
    \mathbf{p}_{\text{new}} = \frac{\mathbf{p}_i + \mathbf{p}_j}{2}
    \]
    This operation was applied across all current points in \( \mathbf{P} \), generating new samples that densely populate boundary regions.\end{enumerate}

The labelling process (or additional experiments) can be carried out until the user is satisfied with the number of labels in the dataset. Experiments shown in Table ~\ref{tab:binary_random} show that high accuracy can be achieved with only 50 images in the training set. 
This method, illustrated in row b) of Figure \ref{fig:labelsize-experiment-comparison}, resulted in significantly improved representation of the domain wall region across all training set sizes. The sampling regime can also be observed in Figure \ref{fig:binary_search_algorithm} and the algorithm can be found in the Appendix. 

\begin{figure}[htbp]
    \centering
    \includegraphics[width=8.3cm]{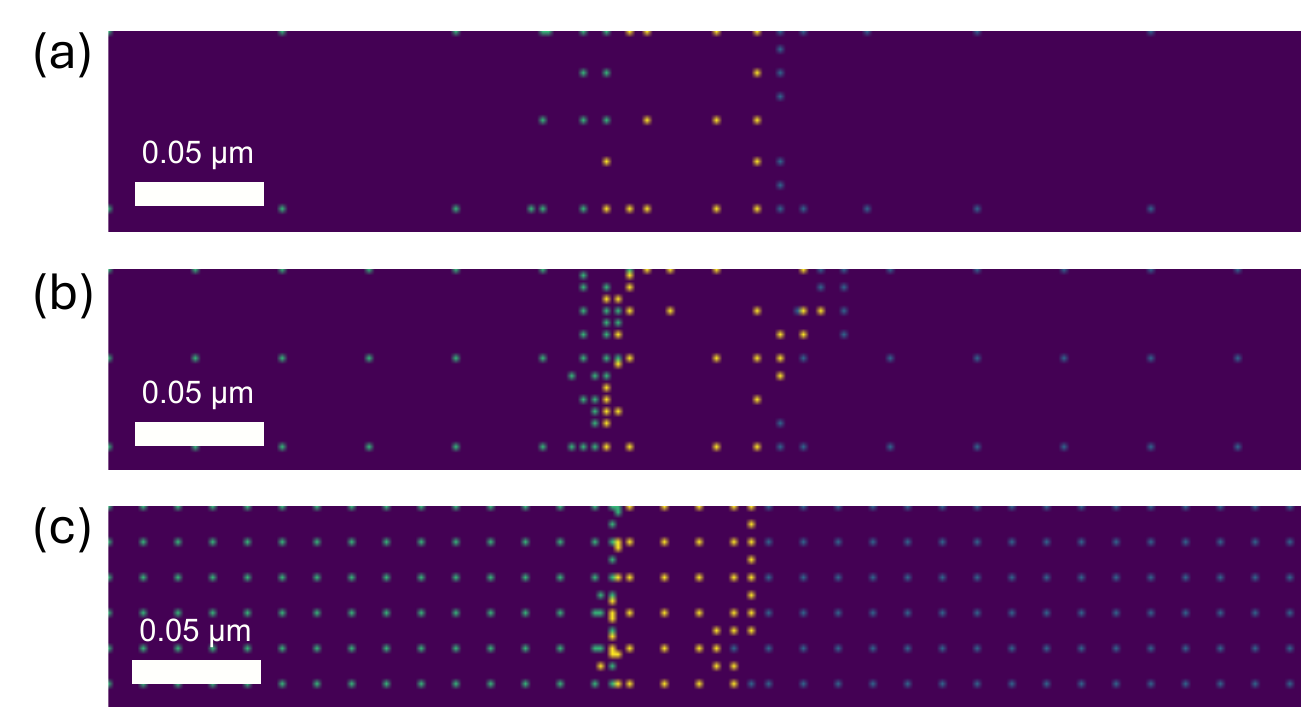}
    \caption{Sampling patterns from real space ($N_{x}, N_{y}$) for the 50 (a), 100 (b) and 250 (c) labelled image datasets withe the adaptive sampling algorithm. The orange dots correspond to the images in the real space that were labelled as vertically polarised, pink dots labelled as horizontally polarised and light blue dots as no polarization observed.}
    \label{fig:binary_search_algorithm}
\end{figure}

A separate test dataset of 1,000 diffraction images was curated, ensuring that none of these images appeared in any of the training datasets used for the models in Figure~\ref{fig:labelsize-experiment-comparison}. All images were hand-labelled, and the accuracy of each model on this test set is reported in Table~\ref{tab:binary_random}. The advantage of the binary search algorithm is evident in the modest yet meaningful increase in accuracy, particularly in the improved classification of domain boundary images, as shown in Figure~\ref{fig:labelsize-experiment-comparison}. Interestingly, increasing the size of the training dataset beyond 50 images does not result in a significant improvement in testing accuracy, highlighting the efficiency of the algorithm with relatively small training sets. Attempts to reduce the training set size below 50 led to poor performance, as the required step size became too large and too few boundary images were included in the search. 

\begin{table}[H]  
\centering
\begin{tabular}{|c|c|c|}
\hline
\textbf{Training Data Size} & \textbf{Binary algorithm accuracy} & \textbf{Random algorithm accuracy} \\
\hline
\textbf{50}  & 96.9\% & 95.0\% \\
\textbf{100} & 96.3\% & 95.7\% \\
\textbf{250} & 96.2\% & 94.2\% \\
\hline
\end{tabular}
\caption{Accuracy comparison for domain polarization assignment on test dataset between binary and random regime models at different training set sizes.}
\label{tab:binary_random}
\end{table}

\section{Methods}
\subsection{Specimen Preparation and 4D-STEM Data Acquisition}

Electron transparent cross-sections of single crystal \ce{Cu3B7O13Cl} were prepared for STEM using a dual-beam focused ion beam (FIB) integrated SEM (Thermo-Fisher Scientific FEI Helios G4 CX). The specimens were mounted onto a DENSsolutions in-situ heating and biasing MEMS chip. Thinning of the samples was done with decreasing accelerating voltage and electron beam current.

All 4D-STEM datasets were acquired on a probe-corrected Thermo Fisher Spectra 300 microscope operating at 300 kV, equipped with a Gatan Continuum energy filter and K3 direct electron detector. The K3 detector was operated in electron counting mode. The convergence angle of the scan was 0.8 mrad, and a 4 × 4 sub-scan was used without drift correction. The scan size was 206 × 34 pixels in real space with a pixel time of 0.02 s. Diffraction patterns were recorded with 512 × 512 pixels.

\subsection{Data Pre-Processing}

The \texttt{py4DSTEM} package \cite{savitzkyPy4DSTEMSoftwarePackage2021} was used to process the Gatan 4D-STEM \texttt{.dm4} files and convert them into a \texttt{.png} format and were then compressed to a \(256 \times 256\) pixel size to improve model training speeds and inference times. 
%
%
The next step required the manual labelling of sampled random images from the 4D-STEM scan. The images were labelled as either horizontally polarized, vertically polarized, or with no observable polarization. This labelled dataset was then used to train the CNN model with a 80:20 train/test split for every experiment, where, to avoid overfitting, the model checkpoint with the lowest validation loss was used at inference and evaluation time. 

\subsection{Model Training}

The model was created and ran using the \texttt{PyTorch} software package \cite{NEURIPS2019_9015}. Training of the model was run on a single Nvidia RTX A6000 Ada Generation GPU for 100 epochs on the full 1000-image training set. Training completed in approximately 6 minutes for each experimental run. Due to the small size of the CNN model (2.1 million parameters), the model was also able to complete inference on all \( 7,004 \) images in the 4D-STEM dataset in 11 seconds.
The AdamW optimizer \cite{loshchilovDecoupledWeightDecay2019} was used for both models with a learning rate of $4\times10^{-4}$, weight decay of $2\times10^{-3}$, momentum of 0.9 and betas of 0.9 and 0.999. A plateau learning rate scheduler was used with a patience of 2, a minimum learning rate of $3\times10^{-5}$ and a maximum learning rate of $5\times10^{-5}$.

\subsection*{Density Functional Theory}

First-principles calculations were performed within DFT as implemented in the Vienna Ab Initio Software Package (VASP)~\cite{VASP1}. The projector-augmented wave method \cite{PAW1,PAW2} was employed, with the following valence configurations: O (2s$^2$2p$^4$), Cl (3s$^2$3p$^5$), B (2s$^2$2p$^1$), and Cu (3p$^6$3d$^{10}$4s$^1$). Exchange-correlation effects were treated using the PBE functional. Spin-polarized calculations were carried out with an on-site Hubbard $U$ of 4 eV applied to Cu $d$ states. This value was chosen by comparing the calculated magnetic properties (magnetic moments and relative energies of magnetic orders) for a range of $U$ values to the metaGGA r2SCAN functional, which is known to have improved accuracy for magnetic materials. For all calculations, we used a plane-wave cut-off energy of 600 eV. The Brillouin zone was sampled with a $4\times4\times4$ and a $3\times3\times2$ Monkhorst-Pack $k$-point grid for the cubic and orthorhombic structures, respectively. 
The electronic convergence criterion is set to $10^{-7}$ eV and the force convergence criterion is set to 0.002 eV/\AA{}, finding good agreement with experimentally reported lattice constants. We do not include spin-orbit coupling in these calculations as it was found to have a negligible impact on the structural phases.

\section{Conclusion}

In summary, we demonstrate that CNNs enable automated identification and segmentation of polarization domains in ferroelectric materials from 4D-STEM diffraction data. This deep learning approach overcomes the limitations of conventional manual analysis, substantially improving both the speed and accuracy of domain mapping. The trained model reliably classifies polarization directions from diffraction features and produces real-space segmentation maps consistent with traditional interpretations.

Beyond domain mapping, our framework highlights the broader potential of machine learning in electron microscopy for the rapid and scalable analysis of large, information-rich 4D-STEM datasets. The ability of the CNN to achieve high accuracy with a minimal training set suggests general applicability to other material systems where subtle structural variations are critical. Furthermore, the labelling algorithm developed here provides a principled means of sampling diffraction data, enabling accurate classification even at domain walls.

Future efforts will extend this methodology to capture dynamic polarization responses under external stimuli and to integrate real-time inference for in situ studies. Together, these advances establish a pathway toward accelerating quantitative analysis and discovery in ferroic materials and related complex systems.

\section{Declarations}
\subsection{Data \& Code Availability}
A repository with the open-source code is available at: \url{https://github.com/FinHardy/ghost-hunter}. 

\subsection{Acknowledgments}
This work was made possible by the EPSRC Cryo-Enabled Multi-microscopy for Nanoscale Analysis in the Engineering and Physical Sciences EP/V007661/1. F. H., M.P., Y.L., and M.S.C. acknowledge funding from the Royal Society Tata University Research Fellowship (URF\textbackslash R1\textbackslash 201318) and Royal Society Enhancement Award RF\textbackslash ERE\textbackslash210200EM1. M.S.C. acknowledges funding that supported this work from ERC CoG DISCO grant 101171966 and EPSRC grant EP/V001914/1. G.T. acknowledges funding from the EPSRC Centre for Doctoral Training in the Advanced Characterisation of Materials (CDTACM)(EP/S023259/1) and Cameca Ltd. for funding their PhD studentship.
This work was funded in part by the U.S. Department of Energy, Office of Science, Office of Basic Energy Sciences, Materials Sciences and Engineering Division under Contract No. DE-AC02-05-CH11231 within the Theory of Materials program (first-principles calculations). Computational resources were provided by the National Energy Research Scientific Computing Center and the Molecular Foundry, DOE Office of Science User Facilities supported by the Office of Science, U.S. Department of Energy under Contract No. DEAC02-05CH11231. The work performed at the Molecular Foundry was supported by the Office of Science, Office of Basic Energy Sciences, of the U.S. Department of Energy under the same contract. 


\bibliography{main}

\subsection{Appendix}

\begin{algorithm}[H]
\caption{Targeted Sampling Strategy for Domain Boundary Representation}
\begin{algorithmic}[1]
\State \textbf{Input:} Image dimensions $(X, Y)$, step size $s$, label map $L$
\State \textbf{Output:} Sampled point set $\mathbf{P}_{\text{aug}}$

\State \textit{// Step 1: Coarse Sampling}
\State Initialize $\mathbf{P} \gets \emptyset$
\For{$i = 0$ to $\lfloor X / s \rfloor$}
    \For{$j = 0$ to $\lfloor Y / s \rfloor$}
        \State $\mathbf{p}_i \gets (i \cdot s, j \cdot s)$
        \State $\mathbf{P} \gets \mathbf{P} \cup \{ \mathbf{p}_i \}$
    \EndFor
\EndFor

\State \textit{// Step 2: Adaptive Sampling near Label Boundaries}
\State Initialize $\mathbf{P}_{\text{new}} \gets \emptyset$
\ForAll{$\mathbf{p}_i \in \mathbf{P}$}
    \ForAll{8 connected neighbours $\mathbf{p}_j$ of $\mathbf{p}_i$}
        \If{$ \text{label of }\mathbf{p}_j \neq  \text{label of } \mathbf{p}_i$}
            \State $\mathbf{p}_{\text{mid}} \gets \frac{\mathbf{p}_i + \mathbf{p}_j}{2}$
            \State $\mathbf{P}_{\text{new}} \gets \mathbf{P}_{\text{new}} \cup \{ \mathbf{p}_{\text{mid}} \}$
            \State \textbf{break}
        \EndIf
    \EndFor
\EndFor

\State $\mathbf{P}_{\text{aug}} \gets \mathbf{P} \cup \mathbf{P}_{\text{new}}$
\State \Return $\mathbf{P}_{\text{aug}}$
\end{algorithmic}
\label{alg:sampling-algorithm}
\end{algorithm}

\end{document}